\documentclass[floats,floatfix,showpacs,amssymb,prd,twocolumn,superscriptaddress,nofootinbib,nolongbibliography,reprint]{revtex4-2}

\usepackage{amssymb,amsmath,verbatim,mathtools,needspace,enumitem,etoolbox,graphicx,physics,microtype,afterpage,bigints,gensymb,tabularx,soul}
\usepackage{bbm}
\usepackage[dvipsnames, usenames]{xcolor}
\definecolor{linkcolor}{rgb}{0.0,0.3,0.5}
\definecolor{dodgerblue}{HTML}{1E90FF}
\usepackage[unicode, colorlinks=true, linkcolor=linkcolor, citecolor=linkcolor, filecolor=linkcolor,urlcolor=linkcolor, pdfusetitle]{hyperref}
\usepackage[all]{hypcap}
\usepackage[T1]{fontenc}
\usepackage[utf8]{inputenc}
\usepackage{orcidlink}
\usepackage{tensor}

\makeatletter
\newcommand*{\balancecolsandclearpage}{\close@column@grid \cleardoublepage \twocolumngrid}
\makeatother

\newcommand{\bham}{\affiliation{School of Physics and Astronomy \& Institute for Gravitational Wave Astronomy, University of Birmingham, \\ Birmingham, B15 2TT, United Kingdom}}
\newcommand{\milan}{\affiliation{Dipartimento di Fisica ``G. Occhialini'', Universit\'a degli Studi di Milano-Bicocca, Piazza della Scienza 3, 20126 Milano, Italy}}
\newcommand{\infn}{\affiliation{INFN, Sezione di Milano-Bicocca, Piazza della Scienza 3, 20126 Milano, Italy}}
\newcommand{\jhu}{\affiliation{William H. Miller III Department of Physics and Astronomy, Johns Hopkins University,
3400 N. Charles Street, Baltimore, Maryland, 21218, USA}}

\begin{document}

\title{Simulation-based inference of black hole ringdowns in the time domain}

\author{Costantino Pacilio$\,$\orcidlink{0000-0002-8140-4992}}
\email{costantino.pacilio@unimib.it}
\milan \infn

\author{Swetha Bhagwat$\,$\orcidlink{0000-0003-4700-5274}$\,$}
\bham

\author{Roberto Cotesta}
\jhu

\date{\today}

\begin{abstract}
Gravitational waves emitted by a ringing black hole allow us to perform precision tests of general relativity in the strong field regime. With improvements to our current gravitational wave detectors and upcoming next-generation detectors, developing likelihood-free parameter inference infrastructure is critical as we will face complications like nonstandard noise properties, partial data and incomplete signal modeling that may not allow for an analytically tractable likelihood function. In this work, we present a proof-of-concept strategy to perform likelihood-free Bayesian inference on ringdown gravitational waves using simulation based inference. Specifically, our method is based on truncated sequential neural posterior estimation, which trains a neural density estimator of the posterior for a specific observed data segment. We setup the ringdown parameter estimation directly in the time domain. We show that the parameter estimation results obtained using our trained networks are in agreement with well-established Markov-chain methods for simulated injections as well as  analysis on real detector data corresponding to GW150914. Additionally, to assess our approach's internal consistency, we show that the density estimators pass a Bayesian coverage test.
\end{abstract}
\maketitle
\section{Introduction}
The detection of gravitational waves (GWs) from binary black-hole (BBH) mergers \cite{LIGOScientific:2016aoc} enables the possibility to perform novel tests of general relativity (GR) in the strong-field regime \cite{Yunes:2016jcc,LIGOScientific:2020tif,LIGOScientific:2021sio}. GR predicts that stationary black holes (BHs) are described by a strikingly simple two-parameter family metric known as the Kerr solution \cite{Kerr:1963ud}, a result that is commonly known as the ``no-hair theorem'' \cite{Israel:1967za,Carter:1971zc}. This is particularly relevant to the final phase of a BBH evolution, where the two black holes have merged to form a perturbed remnant that settles down to a stationary Kerr BH \cite{Teukolsky:2014vca}.

Perturbed BHs shed away multiple moments by emitting GWs \cite{Vishveshwara:1970zz} with characteristic complex oscillation frequencies called quasinormal modes (QNMs) \cite{Kokkotas:1999bd,Berti:2009kk}. In GR, the oscillation frequency spectrum can be predicted from the mass and spin of the remnant BH \cite{Teukolsky:1972my,Berti:2009kk} in accordance with the no-hair theorem. Detecting multiple QNMs  from the observed signal, also known as black-hole spectroscopy \cite{Berti:2005ys,Berti:2007zu}, allows us to test deviations from the BH geometry and in the perturbative equation of motion in GR as well as the boundary conditions imposed to solve the perturbative equation of motion \cite{Dreyer:2003bv,Barausse:2008xv}.

Detecting multiple QNMs unambiguously with present GW detectors has been challenging thus far due to the low signal-to-noise ratio (SNR) in the ringdown portion of the signal  \cite{LIGOScientific:2020tif,LIGOScientific:2021sio,Berti:2007zu} (but see \cite{Capano:2021etf,Siegel:2023lxl} for recent hints of multiple QNMs detections from GW data). On the other hand, it is expected that multiple ringdowns at high SNRs will be observed with next-generation GW detectors such as LISA \cite{2017arXiv170200786A}, Einstein Telescope \cite{Punturo:2010zz}, and Cosmic Explorer \cite{LIGOScientific:2016wof}.

Next-generation detectors are expected to routinely observe ringdown SNRs as high as $\mathcal{O}(100)$, with exceptionally loud events having SNRs greater than $1000$ \cite{Gossan:2011ha,Berti:2016lat,Cabero:2019zyt,Bhagwat:2021kwv,Bhagwat:2023jwv}. Such observations will enable to detect several superimposed QNMs and measure deviations from GR at subpercent level \cite{Bhagwat:2021kwv,Bhagwat:2023jwv}. These will require more powerful Bayesian techniques to both isolate the signal from the background noise and to handle nonstandard noise properties like nonstationarities and non-Gaussianities \cite{Kumar:2022tto,Baghi:2021tfd,Spadaro:2023muy}, and gaps in data as well as systematics in modeling the signal \cite{Toubiana:2023cwr}. Hence, exploring alternative inference methods is of fundamental interest. 

The aim of this paper is to propose a likelihood-free method for the parameter estimation of BH ringdowns, based on simulation-based inference (SBI) \cite{2020PNAS..11730055C}. SBI methods have been previously applied to GW inference of a full inspiral-merger-ringdown signals, starting with the pioneering works of \cite{Chua:2019wwt,Gabbard:2019rde,Green:2020hst,Green:2020dnx} and followed by manifold improvements \cite{Dax:2021tsq,Dax:2022pxd,Wildberger:2022agw,Bhardwaj:2023xph}. All these studies show that SBI reaches an accuracy comparable with traditional Markov-chain methods  for a full inspiral-merger-ringdown waveform \cite{Veitch:2014wba,Biwer:2018osg,Ashton:2021anp,Ashton:2022grj} and also allow for a considerable speedup in the analysis.

In addition, SBI is highly efficient in estimating automatically marginalized posterior densities \cite{Bhardwaj:2023xph}. This is crucial in the parameter estimation of BHs in the ringdown phase when the observable signal includes a very large number of QNMs but, in practice, one is only interested is a small subset of well-resolved modes for testing GR purposes \cite{London:2014cma,London:2018nxs,Baibhav:2023clw,Cheung:2023vki}.

In this work we introduce a time domain implementation of likelihood-free inference of BH ringdowns using SBI. In contrast to frequency domain approaches presented in a recent study \cite{Crisostomi:2023tle}, this technique facilitates noise simulation and requires fewer training samples.
Likelihood-free ringdown inference in the time domain was first explored in \cite{Bhagwat:2021kfa} based on the technique of conditional variational autoencoders \cite{2013arXiv1312.6114K,2016arXiv160605908D}. Here, we demonstrate for the first time that SBI matches the performance of well-established Markov-chain methods for several simulated signal injections \cite{Carullo:2019flw,Isi:2021iql}. We also provide a real-data application of estimating the posterior mass and spin of the final BH remnant from GW150914 \cite{LIGOScientific:2016aoc}, which agrees well with previous Bayesian analyses \cite{LIGOScientific:2016lio}.

\section{Sequential neural posterior estimation}
\label{sec:snpe}
We use sequential neural posterior estimation (SNPE) to estimate posterior densities. SNPE was first introduced in \cite{papamakarios2016fast} and further expanded in \cite{greenberg2019automatic,deistler2022truncated}. The aim of SNPE is to train an approximate density estimator $q_{\boldsymbol{\phi}}(\boldsymbol{\theta}|\boldsymbol{x})$ of the true posterior density $p(\boldsymbol{\theta}|\boldsymbol{x})$. Here, $\boldsymbol{x}$ is the data segment and $\boldsymbol{\theta}$ is the set of model parameters to be estimated; in particular, $\boldsymbol{x}$ is the sum of a deterministic component $\boldsymbol{h}(\boldsymbol{\theta})$ (hereafter identified with the strain of the GW projected on the detector) and a stochastic component $\boldsymbol{n}$ (the noise of the detector), $\boldsymbol{x}(\boldsymbol{\theta})=\boldsymbol{h}(\boldsymbol{\theta})+\boldsymbol{n}$. It is assumed that one can generate $\boldsymbol{h}(\boldsymbol{\theta})$ though a numerical process, so as to prepare a training set of ordered couples $\{\boldsymbol{\theta}_i,\boldsymbol{x}_i\}$ to train the density estimator. 

The training process maximises the likelihood $\prod_i q_{\boldsymbol{\phi}}(\boldsymbol{\theta}_i|\boldsymbol{x}_i)$ with respect to~the model parameters $\boldsymbol{\phi}$. As shown in \cite{papamakarios2016fast}, the estimator converges to the true posterior density asymptotically with the dimension of the training set.
A particular class of estimators called normalizing flows \cite{papamakarios2021normalizing} has become increasingly popular for SNPE. In the following, we model the estimator as a neural spline flow (NSF), a spline-based neural normalizing flow \cite{2019arXiv190604032D,Green:2020dnx}. We use the implementation from the \texttt{sbi} package \cite{tejero-cantero2020sbi} --- see also Appendix \ref{sec:technical} for implementational details.

Once trained, the model $q_{\boldsymbol{\phi}}$ can be sampled from in a fraction of a second, returning fast posterior samples $\{\boldsymbol{\theta}_n\}$. Moreover, differently from Markov-chain methods, the model can also be evaluated and returns fast accurate estimates of $p(\boldsymbol{\theta}|\boldsymbol{x})$. 

The word ``sequential'' refers to the fact that the approximate density estimator $q_{\boldsymbol{\phi}}(\boldsymbol{\theta}|\boldsymbol{x})$ targets a particular observation $\boldsymbol{x}_o$, as opposed to an amortized NPE which targets the entire prior volume of the training set. Specifically, sequential training proceeds through a series of adaptive steps, or rounds: at each round, new training samples are drawn from a proposal distribution $\tilde p(\boldsymbol{\theta})$, which is initially set to the prior $p(\boldsymbol{\theta})$ and then it shrinks adaptively at each round to prioritize samples with high posterior density $q_{\boldsymbol{\phi}}(\boldsymbol{\theta}|\boldsymbol{x})$ conditioned to $\boldsymbol{x}=\boldsymbol{x}_o$.
While the original implementation of SNPE in \cite{papamakarios2016fast} proposes to draw the new samples directly from the posterior of the previous round, we opt for a refined version called truncated SNPE (TSNPE) \cite{deistler2022truncated}, as we observed it to give more stable results. We briefly summarize it here:
\begin{enumerate}
    \item During the first round $r=1$, the training samples $\{\boldsymbol{\theta}_i,\boldsymbol{x}_i\}_{r=1}$ are drawn from the prior $p(\boldsymbol{\theta})$ and the model is optimized with respect to~$\{\boldsymbol{\theta}_i,\boldsymbol{x}_i\}_{r=1}$.
    \item During the $k$-th round ($k>1$), new training samples are still drawn from the prior $p(\boldsymbol{\theta})$, but they are rejected if they fall outside the $1-\epsilon$ highest posterior density (HPD) region of $q_{\boldsymbol{\phi}}(\boldsymbol{\theta}|\boldsymbol{x})$ from the previous round, evaluated at $\boldsymbol{x}=\boldsymbol{x}_o$. The model from the previous round $r=k-1$ is then further optimized with respect to the collection of samples from all rounds $\{\boldsymbol{\theta}_i,\boldsymbol{x}_i\}_{r=1}\cup\dots\cup\{\boldsymbol{\theta}_i,\boldsymbol{x}_i\}_{r=k}$.
    \item The training proceeds until a stopping criterion is met.
\end{enumerate}

Here, $\epsilon$ is an hyperparameter that is fixed in advance and plays a crucial role: a large $\epsilon$ leads to undercovered posteriors due to large truncation of the original prior volume, while a small $\epsilon$ does not adapt the prior enough to advance the training. Reference \cite{deistler2022truncated} reports that the range $10^{-5}<\epsilon<10^{-2}$ is working for a variety of benchmarking problems. In this work, we fix $\epsilon=10^{-4}$ as we found it to give the best results.

To avoid overfitting, we stop the training when the truncated volume at the current round encompasses more than $80\%$ of the truncated volume at the previous round. Moreover, at rounds $r>1$, we also discard simulations from the first round since they are less informative.

More details about the data preparation and the density estimator are provided in Appendix~\ref{sec:technical}.

\section{Ringdown waveform}
\label{sec:ringdown}
The GW strain that we use for the inference has the form
\begin{equation}
    \label{eq:strain:1}
    \boldsymbol{h}(\boldsymbol{\theta})=F_+h_+ + F_\times h_\times
\end{equation}
where $F_+$ and $F_\times$ are the pattern functions of the detector  \cite{Jaranowski:1998qm}, depending on the sky position and relative orientation of the source, of the polarization of the waveform and of the starting time of the ringdown. The plus- and cross- waveform components are expressed as superpositions of damped sinusoids as

\begin{subequations}
\begin{align}    
    \label{eq:ringdown:1}
    &h_+=\sum_{l,m,n}\mathcal{A}_{lmn}e^{-\frac{(t-t_{\rm start})}{\tau_{lmn}}}\cos(\Phi_{lmn})Y_{lm}^+(\iota)\\
    &h_\times=\sum_{l,m,n}\mathcal{A}_{lmn}e^{-\frac{(t-t_{\rm start})}{\tau_{lmn}}}\sin(\Phi_{lmn})Y_{lm}^\times(\iota)
\end{align}
\end{subequations}
where $l\geq2$, $|m|\leq l$ and $n\geq0$, and
\begin{equation}
    \Phi_{lmn}\equiv2\pi f_{lmn}(t-t_{\rm start})+\phi_{lmn}\,.
\end{equation}
The discrete indices $(l,m,n)$ label the (complex) QNMs $\tilde\omega_{lmn}=2\pi f_{lmn}+i/\tau_{lmn}$ of the remnant black hole, with $f_{lmn}$ and $\tau_{lmn}$ being the $(l,m,n)$ frequency and damping time, respectively. While the QNM indices $(l,m,n)$ span a countably infinite set, numerical simulations show that only a finite subset is significantly excited in the aftermath of a binary black-hole merger \cite{Berti:2005ys,Berti:2007zu,London:2014cma}. The amplitudes $\mathcal{A}_{lmn}$ quantify the extent to which different QNMs are excited \cite{Berti:2007zu,Gossan:2011ha,Kamaretsos:2012bs,London:2014cma,London:2018gaq,JimenezForteza:2020cve,Forteza:2022tgq}.

The expressions \eqref{eq:ringdown:1} assume nonprecessing progenitors, implying the equatorial symmetry
\begin{equation}
    \label{eq:equatorial}
    \mathcal{A}_{l-mn}e^{i\phi_{l-mn}}=(-1)^l\mathcal{A}_{lmn}e^{-i\phi_{lmn}}\,.
\end{equation}
Together with the symmetry $\tilde\omega_{l-mn}=-\tilde\omega_{lmn}^*$ valid for the QNMs of Kerr BHs, it allows us to reabsorb negative-$m$ modes into positive-$m$ ones and to restrict the sums over $m>0$ only.

The plus- and cross- spherical harmonics are functions of the inclination angle $\iota$ defined as 
\begin{equation}
    Y_{lm}^{+,\times}(\iota)=\tensor[_{-2}]{Y}{_{lm}}(\iota,0)\pm(-1)^l\tensor[_{-2}]{Y}{_{l-m}}(\iota,0)
\end{equation}
in terms of the spin-weighted spherical harmonics $\tensor[_{-2}]{Y}{_{lm}}$. Note that we set the azimuthal angle to zero because it is degenerate with the QNM phases. We also take the the spin-weighted spherical harmonics $\tensor[_{-2}]{Y}{_{lm}}$ to approximate the spin-weighted spheroidal harmonics \cite{Berti:2014fga}, $\tensor[_{-2}]{Y}{_{lm}}\approx\tensor[_{-2}]{S}{_{lmn}}$. 

For the parameter estimation, we follow \cite{Isi:2019aib,Cotesta:2022pci,Capano:2021etf} and keep sky position, polarization, starting time and inclination as fixed parameters.

As a consequence, the space of independent model parameters is spanned by $\boldsymbol{\theta}=\{M,\chi\}\cup\{\mathcal{A}_{lmn},\phi_{lmn}\}_{lmn}$ and has $2(1+N_\text{modes})$ dimensions. We assume that GR is valid and we map $\{M_f,\chi_f\}$ into the Kerr QNMs using the fits in \cite{London:2018nxs}.

\section{Simulated injections}
We perform simulated injections into zero noise \cite{Nissanke:2009kt,Rodriguez:2013oaa} to benchmark the ability to recover a known set of injected parameters. We simulate three systems: 
\begin{enumerate}
    \item $\text{Kerr}_{220}$: A system containing only the $(2,2,0)$ mode. This system is designed to benchmark our inference strategy on the simplest problem of a single excited mode.
    \item $\text{Kerr}_{221}$: A system containing the $(2,2,0)$ mode and the $(2,2,1)$ mode, checking the ability to recover the fundamental tone and its first overtone.
    \item $\text{Kerr}_{330}$: A system containing the $(2,2,0)$ and the $(3,3,0)$ mode, checking the ability to recover the fundamental tone and a higher angular mode.
\end{enumerate}

\begin{figure*}[t]
    \centering
    \includegraphics[width=0.3\textwidth]{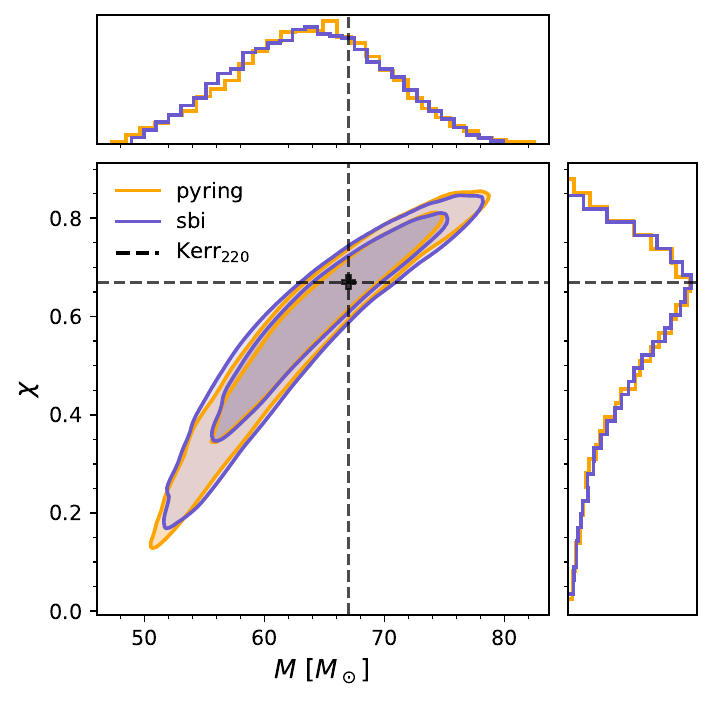}
    \includegraphics[width=0.3\textwidth]{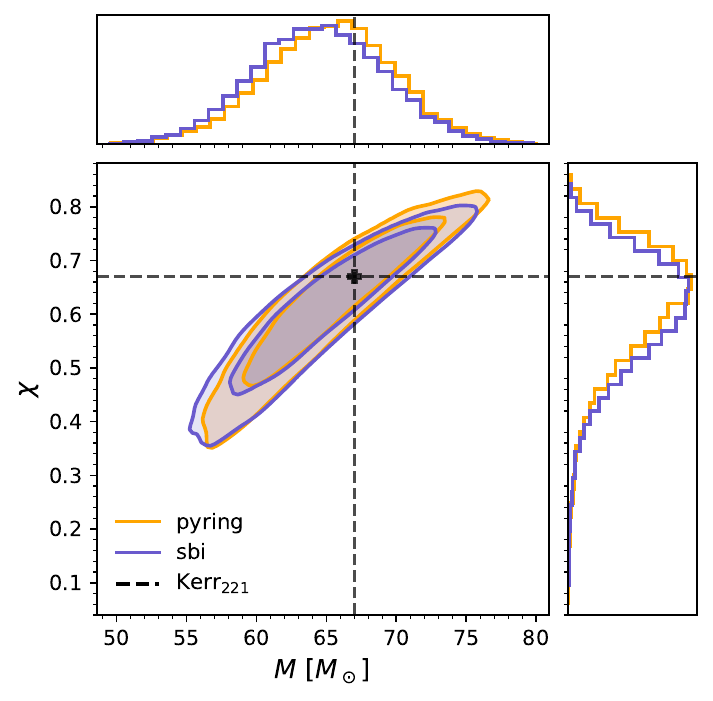}
    \includegraphics[width=0.3\textwidth]{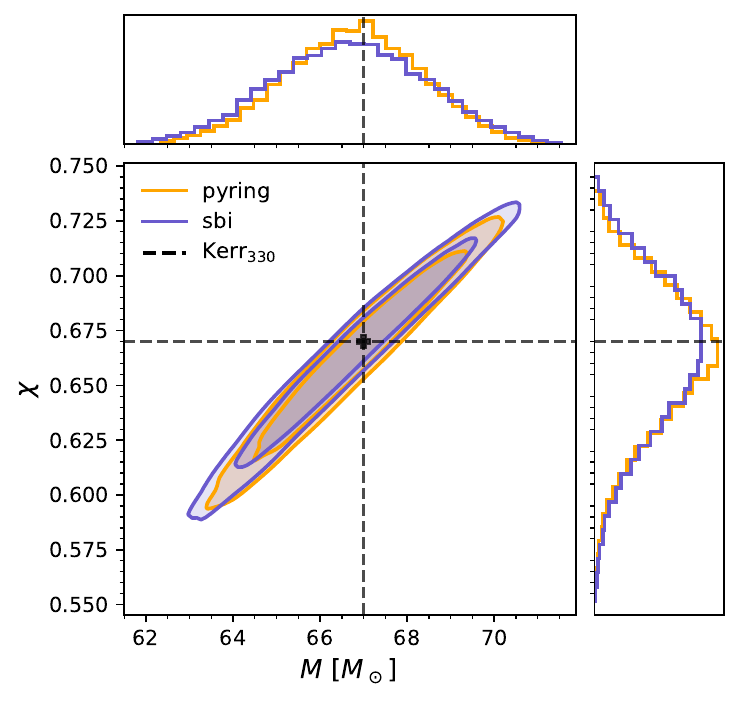}
    \caption{Marginalized posteriors for $(M_f,\chi_f)$ for the three injected systems. The systems $\text{Kerr}_{220}$ and $\text{Kerr}_{221}$ have $\text{SNR}\approx14$ while the system $\text{Kerr}_{330}$ has $\text{SNR}\approx53$, which justifies its narrower posteriors. 2-D contours denote $68\%$ and $90\%$ credible regions. Black dashed lines denote the injected parameters.}
    \label{fig:mchi:inj}
\end{figure*}
In all cases, the mass and spin of the remnant BH are fixed to $M_f=67~M_\odot$ and $\chi_f=0.67$. The other parameters are listed in Table \ref{tab:injections}. For the $\text{Kerr}_{220}$ and the $\text{Kerr}_{221}$ systems, we fix the inclination angle to the GW150914-like value $\iota=\pi$ \cite{Isi:2019aib,Cotesta:2022pci}. Since the spherical harmonics $Y_{330}^{+,\times}$ vanish at $\iota=\pi$, we fix $\iota=\pi/4$ for the $\text{Kerr}_{330}$ system, in order to excite the higher angular mode.

Following \cite{Isi:2019aib}, we fix the right ascension $\alpha$, declination $\delta$ and polarization $\psi$ to the GW150914-like values $\{\alpha,\delta,\psi\}=\{1.95,-1.27,0.82\}$. Similarly, we fix the starting time of the ringdown to the median peak time estimated from the posterior samples of GW150914 \cite{gwtc-1-samples,Cotesta:2022pci}; specifically, we fix the GPS starting time at H1 to be $t_{\rm H1}=1126259462.42323$, while the starting time at L1 is derived from the time delay from H1 using the sky location fixed above.

The SNRs are computed from the noise power spectral density (PSD), estimated from the data segments in the vicinity of the event GW150914 \cite{Isi:2021iql}. The amplitudes of $\text{Kerr}_{220}$ and the $\text{Kerr}_{221}$ systems are adjusted to give a signal-to-noise ratio $\text{SNR}\approx14$, comparable to the SNRs of ringdown events observed so far, e.g., GW150914. On the other hand, $\text{SNR}\approx14$ is too low to allow for a detection of the subdominant $(3,3,0)$ mode: following \cite{Berti:2007zu,Berti:2016lat}, we expect that the higher mode $(3,3,0)$ is resolved from the $(2,2,0)$ starting from a ringdown SNR of $\mathcal{O}(50)$, for nonspinning progenitors with mass ratio $q\approx1.5$. Therefore, we fix the corresponding amplitudes to $\{\mathcal{A}_{220},\mathcal{A}_{330}\}=\{30\times10^{-21},3\times10^{-21}\}$, which results into $\text{SNR}\approx53$.

\renewcommand{\arraystretch}{1.} 
\begin{table}[t]
    \centering
    \begin{tabular}{|c|c|c|c|c|c|c|c|}
         \hline
         \multicolumn{8}{|c|}{$\text{Kerr}_{220}:~(2,2,0)$}\\
         \hline
         $M_f$ & $\chi_f$ & $\mathcal{A}_{220}$ & & $\phi_{220}$ &  & $\iota$ & SNR\\
         $67~M_\odot$ & $0.67$ & $5$ & & $1.047$ & & $\pi$ & 14\\
         \hline
         \multicolumn{8}{c}{}\\
         \hline
         \multicolumn{8}{|c|}{$\text{Kerr}_{221}:~(2,2,0)+(2,2,1)$}\\
         \hline
         $M_f$ & $\chi_f$ & $\mathcal{A}_{220}$ & $\mathcal{A}_{221}$ & $\phi_{220}$ & $\phi_{221}$ & $\iota$ & SNR\\
         $67~M_\odot$ & $0.67$ & $8.92$ & $9.81$ & $1.047$ & $4.19$ & $\pi$ & 14\\
         \hline
         \multicolumn{8}{c}{}\\
         \hline
         \multicolumn{8}{|c|}{$\text{Kerr}_{330}:~(2,2,0)+(3,3,0)$}\\
         \hline
         $M_f$ & $\chi_f$ & $\mathcal{A}_{220}$ & $\mathcal{A}_{330}$ & $\phi_{220}$ & $\phi_{330}$ & $\iota$ & SNR\\
         $67~M_\odot$ & $0.67$ & $30$ & $3$ & $1.047$ & $5.014$ & $\pi/4$ & 53\\
         \hline
    \end{tabular}
    \caption{Parameters of the simulated zero-noise injections. Amplitudes are expressed in units of $10^{-21}$.}
    \label{tab:injections}
\end{table}

\begin{figure}[t]
    \centering
    \includegraphics[width=0.35\textwidth]{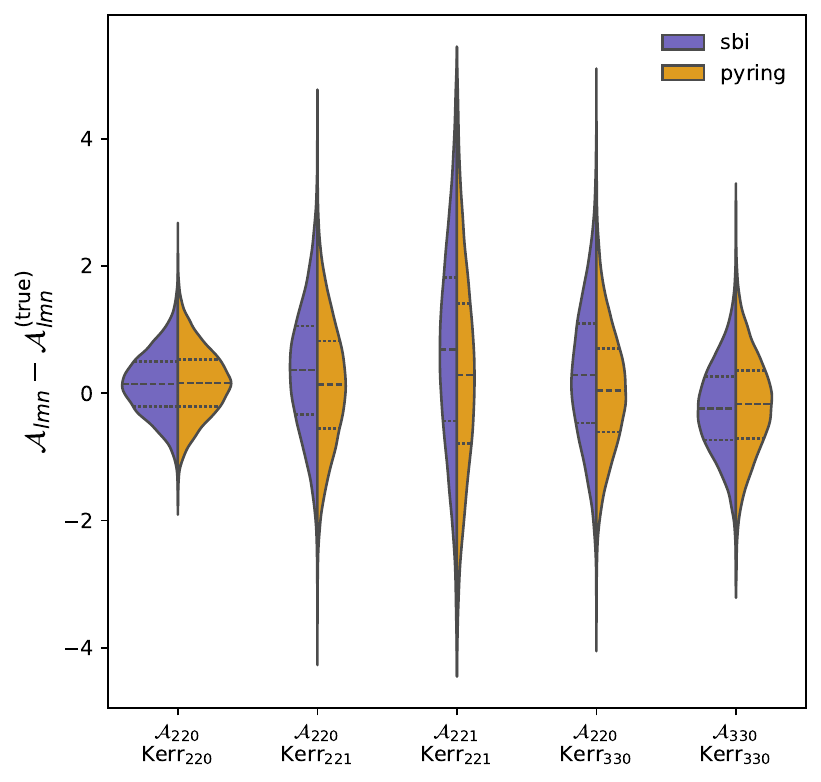}
    \caption{Violin plots of the recovered amplitudes $\mathcal{A}_{lmn}$ for each injected model. We subtract the injected values $\mathcal{A}_{lmn}^{\rm (true)}$ for better visualization. All amplitudes are expressed in units of $10^{-21}$.}
    \label{fig:violin}
\end{figure}

We infer the posteriors from box-uniform priors in the model parameters, within the prior ranges listed in Table \ref{tab:priors}. We benchmark our results against posterior samples obtained with the time-domain inference software \texttt{pyRing} (v2.3.0) \cite{pyRing,Carullo:2019flw,Isi:2019aib,LIGOScientific:2020tif}. 
We sample with the nested sampler \texttt{cpnest} \cite{cpnest}, using 4096 live points and 4094 maximum Markov-chain steps, which typically results in $\sim 20$k posterior samples. For consistency, we extract 20k posterior samples from the trained estimator $q_{\boldsymbol{\phi}}(\boldsymbol{\theta}|\boldsymbol{x}_o)$ when doing the comparisons.

\begin{table}[t]
    \centering
    \begin{tabular}{|c|c|}
        Parameter & Prior range\\
        \hline
         $M_f$ &  $[20,300]~M_\odot$\\
         $\chi_f$ &  $[0,0.99]$ \\
         $\mathcal{A}_{220}$ &  $[0.1,50]\times10^{-21}$ \\
         $\mathcal{A}_{221}$ &  $[0,50]\times10^{-21}$ \\
         $\mathcal{A}_{330}$ &  $[0,50]\times10^{-21}$ \\
         $\phi_{lmn}$ &  $[0,2\pi]$
    \end{tabular}
    \caption{Prior ranges for inferring the model parameters. All priors are box-uniform distributions within the specified ranges.}
    \label{tab:priors}
\end{table}
We train the posterior density estimators on a single GPU A100. Individual training processes take $\{291,324,249\}$ epochs and last $\{81,70,52\}$ mins for the $\{\text{Kerr}_{220},\text{Kerr}_{221},\text{Kerr}_{330}\}$ models respectively. 
Figure \ref{fig:mchi:inj} displays the marginalized posteriors for $(M_f,\chi_f)$ obtained with \texttt{pyRing} and with our SBI approach, showing an excellent agreement between the two methods.
Figure \ref{fig:violin} shows that amplitudes, containing the physical information on the degree of excitation of the QNMs, are recovered consistently with Markov-chain methods for all injected systems.
We also display full corner plots for the $\text{Kerr}_{221}$ and $\text{Kerr}_{330}$ in Appendix \ref{sec:corner:plots}.

As an internal diagnostics, we perform coverage tests \cite{cook2006validation,2018arXiv180406788T,Karchev:2022xyn}. In particular we check that, over a set of simulated injections from the restricted priors $\tilde p(\boldsymbol{\theta})$, the Bayesian credible intervals can be used similar to frequentist confidence regions; equivalently, that the true injected values fall within the $\gamma\%$ credible regions in a fraction $\gamma$ of the injections. This is a necessary condition for the density estimators to provide a consistent parameter estimation. Further details are provided in Appendix \ref{sec:coverage}. Figure \ref{fig:coverage:1} displays the cumulative distribution ${\rm c.d.f.}(\gamma)$ of the Bayesian credible intervals for the model ${\rm Kerr}_{220}$: in order to check consistency with the identity, we evaluate $\text{c.d.f.}(\gamma)$ for 100 catalogs, each consisting of 100 events. We see that ${\rm c.d.f.}(\gamma)$ has support well within the expected $90\%$ credible interval around the diagonal line, represented by the gray region \cite{LIGOScientific:2021sio}\footnote{Note that such tests are difficult with ordinary Markov-chain methods, in which typically the densities are sampled from but not evaluated exactly, often restricting the scope of coverage diagnostics to the marginalized 1-d distributions \cite{Romero-Shaw:2020owr}.}.
We obtain similar consistency results for the ${\rm Kerr}_{221}$ and ${\rm Kerr}_{330}$ models, as displayed in Appendix \ref{sec:coverage}.

\begin{figure}[t]
    \centering
    \includegraphics[width=0.35\textwidth]{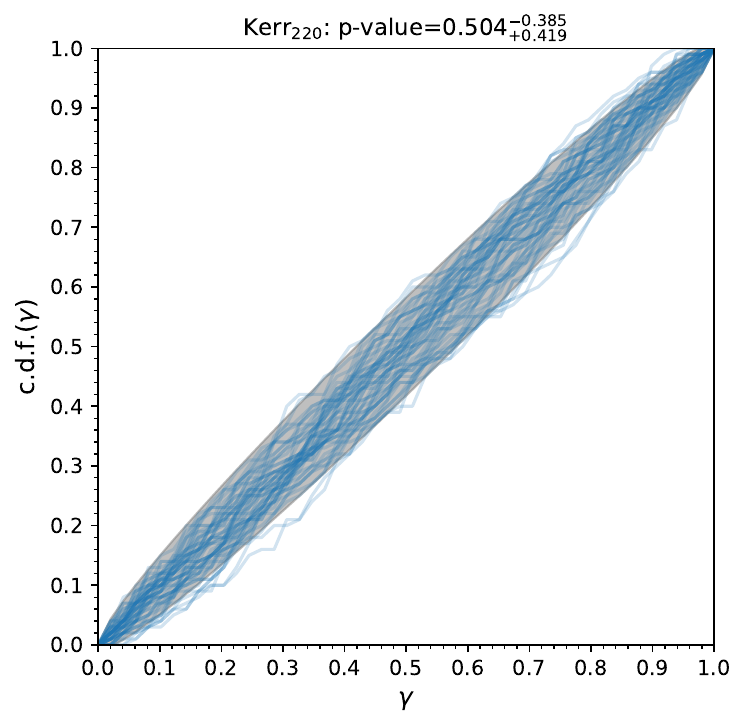}
    \caption{Cumulative distribution of the coverage $\gamma(\boldsymbol{\theta}_*,\boldsymbol{x}_*)$ for the injected model ${\rm Kerr}_{220}$. Each of the 100 blue lines corresponds to the c.d.f.~of $\gamma$ from $N_s=100$ draws of $\boldsymbol{\theta}_*$. The shaded gray area denote the $90\%$ uncertainty over $\text{c.d.f.}(\gamma)$. We also quote the median and $90\%$ confidence bounds of the KS test p-values across all draws.}
    \label{fig:coverage:1}
\end{figure}

Our results show that the time-domain SBI inference for ringdown returns posterior parameter predictions that are internally consistent and it reproduces the results obtained by the established Markov-chain methods.

\section{Inference of real data}
\begin{figure}[t]
    \centering
    \includegraphics[width=0.43\textwidth]{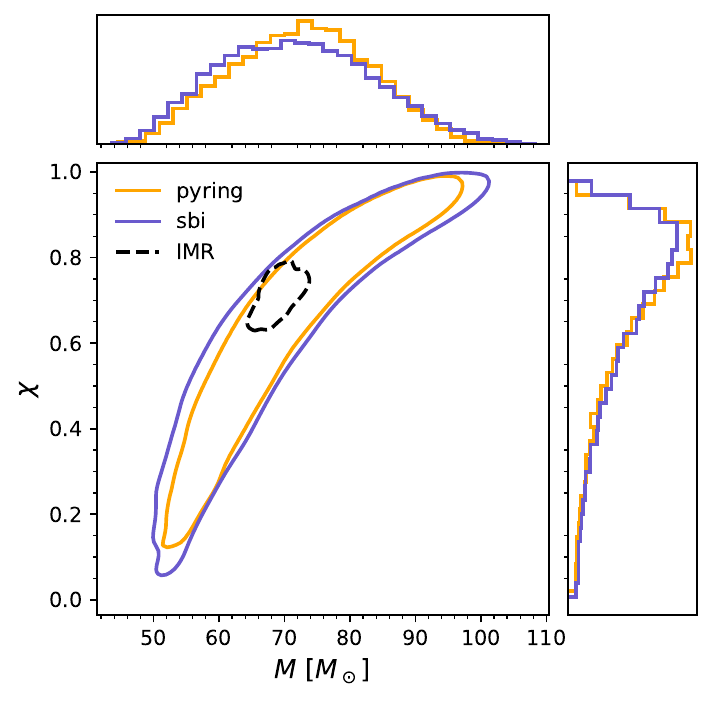}
    \caption{Remnant properties of GW150914, obtained by analyzing the publicly released data from LVK \cite{gwtc-1-documentation}. Solid lines denote the $90\%$ credible regions reconstructed with our SBI implementation (blue) and with the \texttt{pyRing} analysis (orange). We also display the final mass and spin obtained from the full IMR posterior samples \cite{gwtc-1-samples} through the phenomenological remnant fits in \cite{Barausse:2012qz,Hofmann:2016yih}.}
    \label{fig:mchi:real}
\end{figure}

We perform a Bayesian parameter estimation on the real event GW150914 \cite{LIGOScientific:2016aoc}. A full inspiral-merger-ringdown (IMR) analysis identified it as a binary BH with detector-frame total mass $M=72^{+4}_{-3}~M_\odot$ and binary mass ratio $q<1.42$ at $90\%$ credibility, while being compatible with vanishing progenitor spins \cite{gwtc-1-samples}. In order to reconstruct the remnant properties, we analyze the event at $t_{\rm H1}+3~{\rm ms}$, corresponding to $\sim 10 M$ after the merger ($t_{\rm start}\sim t_{\rm peak}+10M$ was found to mark the onset of the ringdown stage from the analysis of nearly equal mass NR simulations \cite{London:2014cma,Bhagwat:2017tkm}). This choice results in a ringdown-only ${\rm SNR}\sim 8.5$ \cite{LIGOScientific:2016lio}. As in \cite{Isi:2019aib}, we fix $\{\alpha,\delta,\phi\}=\{1.95,-1.27,0.82\}$ and $\iota=\pi$. We assume a model containing only the fundamental mode $(2,2,0)$, which dominates the late ringdown stage for nearly equal-mass nonspinning binaries \cite{Kamaretsos:2011um,Kamaretsos:2012bs,Gossan:2011ha,London:2014cma}. 
The training takes 282 epochs and lasts 91 minutes.
Figure \ref{fig:mchi:real} shows the $90\%$ credible regions of the reconstructed final mass and final spin using our SBI implementation, compared to the estimates from the \texttt{pyRing} software. The two results are in good agreement with each other. In the plot, we also display the final mass and spin deduced from the IMR posterior samples \cite{gwtc-1-samples} using the phenomenological fits in \cite{Barausse:2012qz,Hofmann:2016yih}.

\section{Conclusions and outlook}
\label{sec:conclusion}
Our study shows that simulation-based GW inference of BBH ringdowns gives results that are both internally consistent and compatible with Markov-chain methods. After benchmarking SBI on a control set of injections at zero noise, we applied our methods to the analysis of real GW data from LIGO detectors.
We see that inferences done using our model on the event GW150914 agrees with the inference of the traditional setup. It is important to note that zero-noise injections include several simplifications, but real data present noise and a morphological bias between the template and the underlying signal. The fact that our results are consistent in both cases constitutes a robust demonstration of the feasibility of our approach.

We envisage a number of future studies to further expand our results. First, a detailed investigation of all the promising ringdown events from the GW observations is ongoing. While the significance for detecting a second mode in present ringdown data is still debated \cite{LIGOScientific:2020ufj,LIGOScientific:2020tif,LIGOScientific:2021sio,Cotesta:2022pci,Isi:2022mhy,Crisostomi:2023tle}, there is tentative evidence that GW150914 and GW190521 \cite{LIGOScientific:2020iuh} allow us to measure the presence of a subdominant overtone \cite{Isi:2019aib,Finch:2022ynt} and of a higher angular harmonic \cite{Capano:2021etf,Siegel:2023lxl}, respectively. We seek to confirm or refute these findings with our independent, likelihood-free method. 

A promising application leverages on the ability of SBI to automatically marginalize over nuisance parameters. In particular, when inferring on ringdown signals, one is only interested in the small subset of modes which are better resolved. The presence of multiple modes, which are challenging to resolve individually, is a potential source of confusion noise and therefore it must be modeled. Moreover, an ideal model would allow for a variable number of such modes, in order to fit the residuals with greater flexibility. This problem is especially suited for SBI, as a variable number of background modes can be regarded as a collection of nuisances, and we will present a dedicated exploration in a follow-up work.

Lastly, we use sequential (i.e., nonamortised) NPE because it is convenient for analyzing individual observations, which is the scope of our work. On the other hand, amortized NPE might prove useful when assessing the expected performances of a detector on large population catalogs. Such studies are well timed, given the recent adoption of the LISA mission \cite{2017arXiv170200786A} by the European Space Agency, and the strong efforts to build next-generation ground-based GW detectors \cite{Punturo:2010zz,LIGOScientific:2016wof}. Relevant examples include \cite{Branchesi:2023mws} exploring a battery of diverse science cases, and~\cite{Berti:2016lat,Bhagwat:2021kwv,Bhagwat:2023jwv,Maselli:2023khq,Yi:2024elj} with a focus on BH spectroscopy. Since performing rigorous Bayesian analysis on tens of thousands of sources is computationally challenging, all these works approximate individual posterior densities with Fisher matrices, which are known to introduce systematic biases in the shape of the posterior recovery \cite{Vallisneri:2007ev}. Training an amortized density estimator on the expected noise spectral densities would allow to drop such approximations and to draw much more robust conclusions. 

\section{Acknowledgments}
We are especially grateful to Enrico Barausse and Marco Crisostomi for fruitful exchanges during an earlier stage of this work. CP thanks Elliot Finch for useful discussions.
C.P.~is supported by ERC Starting Grant No.~945155--GWmining, 
Cariplo Foundation Grant No.~2021-0555, MUR PRIN Grant No.~2022-Z9X4XS, 
and the ICSC National Research Centre funded by NextGenerationEU. 
CP acknowledges support from the European Union’s H2020 ERC
Consolidator Grant ``GRavity from Astrophysical to Microscopic
Scales'' (Grant No. GRAMS-815673) during a visit to SISSA.
S.B. is supported by UKRI Stephen Hawking Fellowship No.~EP/W005727.
R.C. is supported by NSF
Grants No. AST-2006538, PHY-2207502, PHY-090003
and PHY-20043, and by NASA Grants No. 20-LPS20-0011
and 21-ATP21-0010.
Computational work was performed on the LEONARDO HPC facility hosted at CINECA, with allocations 
through INFN and Bicocca.
\paragraph*{Software.}
We used the following public software: \texttt{scipy} \cite{2020SciPy-NMeth} and \texttt{pycbc} \cite{Biwer:2018osg} to estimate the PSD and the correlation matrix from noisy time series; \texttt{sbi} \cite{tejero-cantero2020sbi} for simulation based inference; \texttt{pyRing} \cite{pyRing} for Markov-chain ringdown inferences; \texttt{numpy} \cite{harris2020array}, \texttt{matplotlib} \cite{Hunter:2007} and \texttt{seaborn} \cite{Waskom2021} for chain visualizations.
\clearpage
\appendix
\section{Technical details}
\label{sec:technical}
In this Appendix, we provide technical details on the process of data preparation, and on the numerical implementation and training of the neural density estimator.
\subsection{Data preparation}
\label{sec:data}
When performing simulated injections, we sample the strains at $2048$ Hz and truncate them at a duration of $0.1$ s, thus resulting in a data segment of $204$ bins per detector. When analyzing real data, we use the publicly available data from GWOSC \cite{gwtc-1-documentation} at $4096$ Hz and down-sample them to $2048$ Hz. 

We whiten the data directly in the time domain following \cite{Isi:2021iql},
\begin{equation}
    \label{eq:whiten:1}
    h_{\rm white}^a(\boldsymbol{\theta}) = \sum_b\tensor{(L^{-1})}{^a_b}h^b(\boldsymbol{\theta})
\end{equation}
where the matrix $L$ is the (upper-triangular) Cholesky factor of the covariance matrix. Specifically, we first compute the auto-correlation function (ACF) from the inverse-Fourier transform of the PSD, then, we compute the covariance matrix as the Toeplitz matrix of the ACF (see Eq.s~(39)-(44) of \cite{Isi:2021iql}). Noise realizations can be added to the whitened ringdown strain as simply as
\begin{equation}
    \label{eq:strain:2}
    \boldsymbol{x}_{\rm white}(\boldsymbol{\theta}) = \boldsymbol{h}_{\rm white}(\boldsymbol{\theta}) + \mathcal{N}(0,1)\,.
\end{equation}

\subsection{Density estimator}
\label{sec:estimator}
We model the density estimator as a neural spline flow (NSF) \cite{papamakarios2021normalizing,Green:2020dnx} and use the implementation from the \texttt{sbi} package \cite{tejero-cantero2020sbi}. The relevant hyper-parameters are listed in Table \ref{tab:nsf}. In particular, the NSF is a flow of 5 transforms and each hidden layer contains 150 units. 

The NSF does not take $\boldsymbol{x}_{\rm white}$ in input directly, but we apply an embedding network for dimensional reduction: the raw input is a concatenation of $\boldsymbol{x}_{\rm white}$ from two detectors (LIGO-Hanford and LIGO-Livingstone), thus resulting into an input dimension of $204+204$ bins; they are then mapped into 128 bins by a fully connected neural network with two hidden layers consisting of 150 units. The output of the embedding network is fed into the NSF. 

We train and validate the density estimator in batches of 512 samples. During the first round we generate 50k training samples, while at later rounds we generate 100k additional samples. We use the Adam optimizer with learning rate 0.001.

Since neural networks learn better from standardized and/or normalized data \cite{LeCun1998}, we linearly rescale $\boldsymbol{x}_{\rm white}$ to have zero mean and unit variance along the feature dimension \cite{George:2016hay}, and we normalize the parameters $\boldsymbol{\theta}$ between 0 and 1. 

Finally, we vary the noise realizations in \eqref{eq:strain:2} at each training epoch to make the inference resilient to specific noise realizations\footnote{To avoid confusion, we clarify the distinction between training rounds and training epochs. A training round is the training update of the density estimator at each truncation of the prior volume. A training epoch is a single feed-forward and back-propagation step of the training set into the network. Each training round consists of several-to-many training epochs, necessary to optimize the model at the current round.}. Note that the original implementation of \texttt{sbi} does not allow to vary noise realizations between training epochs. We leverage on the flexibility of object-oriented programming to wrap the original SNPE implementation of \texttt{sbi} and we redefine its loss method, so as to resample the noise at each evaluation of the loss\footnote{Such a data-augmentation strategy is feasible, because sampling a standard normal in \eqref{eq:strain:2} is a fast numerical operation and it does not impact on the training time. In cases where noise generation is computationally costly, one would opt for alternative data augmentations, e.g., preparing the training set offline with multiple copies of the same raw strains $\boldsymbol{h}(\boldsymbol{\theta})$ but different noise realizations. See \cite{Crisostomi:2023tle} for an example of the last strategy.}.
\begin{table}[t]
    \centering
    \begin{tabular}{c|c}
        \multicolumn{2}{c}{Neural spline flow}\\
        \hline
         \texttt{num\_blocks} & 2 \\
         \texttt{hidden\_features} & 150  \\
         \texttt{num\_transforms} & 5 \\ 
         \texttt{num\_bins} & 10 \\
         \texttt{batch\_norm} & True\\
        \multicolumn{2}{c}{}\\
        \multicolumn{2}{c}{Embedding FC network}\\
        \hline
        \texttt{input\_dim} & 408\\
        \texttt{num\_hidden\_layers} & 2\\
        \texttt{hidden\_dim} & 150\\
        \texttt{output\_dim} & 128\\
        \multicolumn{2}{c}{}\\
        \multicolumn{2}{c}{Training hyper-parameters}\\
        \hline
        \texttt{num\_simulations} & [50k,100k,\dots]\\
        \texttt{batch\_size} & 512 \\
        \texttt{learning\_rate} & 0.001\\
        \texttt{validation\_fraction} & 0.1\\
        \texttt{trunc\_quantile} $\epsilon$ & $10^{-4}$\\
        \texttt{stopping\_ratio} & 0.8\\
        \texttt{varying\_noise} & True
    \end{tabular}
    \caption{Hyper-parameters for the architecture and training of the the density estimator.}
    \label{tab:nsf}
\end{table}
\section{Coverage Test}
\label{sec:coverage}
Given model parameters $\boldsymbol{\theta}_*$ and a corresponding model realization $\boldsymbol{x}_*=\boldsymbol{h}(\boldsymbol{\theta}_*)+\boldsymbol{n}$, we define the coverage $\gamma(\boldsymbol{\theta}_*,\boldsymbol{x}_*)$ as the approximate posterior probability contained within the highest posterior density (HPD) region which has $\boldsymbol{\theta}_*$ at its boundary,
\begin{equation}
    \label{eq:coverage:1}
    \gamma(\boldsymbol{\theta}_*,\boldsymbol{x}_*) = \int d\boldsymbol{\theta}~q_{\boldsymbol{\phi}}(\boldsymbol{\theta}|\boldsymbol{x}_*)
    \mathbbm{1}\left[q_{\boldsymbol{\phi}}(\boldsymbol{\theta}|\boldsymbol{x}_*)>
    q_{\boldsymbol{\phi}}(\boldsymbol{\theta_*}|\boldsymbol{x}_*)\right]
\end{equation}
where $q_{\boldsymbol{\phi}}$ is the density estimator.

The integral \eqref{eq:coverage:1} can be evaluated efficiently via importance sampling as the expectation value 
\begin{equation}
    \label{eq:coverage:2}
    \gamma(\boldsymbol{\theta}_*,\boldsymbol{x}_*)\approx\mathbb{E}_{
    \boldsymbol{\theta}\sim q_{\boldsymbol{\phi}(\boldsymbol{\theta}|\boldsymbol{x}_*)}}
    \big(\mathbbm{1}\left[q_{\boldsymbol{\phi}}(\boldsymbol{\theta}|\boldsymbol{x}_*)>
    q_{\boldsymbol{\phi}}(\boldsymbol{\theta_*}|\boldsymbol{x}_*)\right]
    \big)
\end{equation}
thanks to the fact that the neural density estimator is fast to sample and to evaluate. 
It can be shown \cite{2018arXiv180406788T,Karchev:2022xyn} that, if the density estimator $q_{\boldsymbol{\phi}}(\boldsymbol{\theta}|\boldsymbol{x})$ approximates the true posterior $p(\boldsymbol{\theta}|\boldsymbol{x})$, then the coverage across the prior $p(\boldsymbol{\theta})$ is distributed uniformly within $[0,1]$. The latter statement is equivalent to $\text{c.d.f.}(\gamma)\equiv\int_0^\gamma d\gamma' p(\gamma')=\gamma$ for each $\gamma\equiv\gamma(\boldsymbol{\theta}_*,\boldsymbol{x}_*)$ and $\boldsymbol{\theta}_*\sim p(\boldsymbol{\theta})$. This is a necessary condition for the density estimator to be a valid inference and we check for it. 

Since we are performing truncated SNPE, we cannot sample the injected parameters from the original prior, $\boldsymbol\theta_*\sim p(\boldsymbol{\theta})$. Indeed, sequential training progressively truncates the prior volume as the training proceeds and the density estimator is only guaranteed to work on the final version of the truncated prior $\tilde p(\boldsymbol{\theta})$. Therefore, in the case of TNSPE, the prior for the Bayesian coverage test must correspond to the final truncated prior, $\boldsymbol\theta_*\sim\tilde p(\boldsymbol{\theta})$.

\begin{figure}[t]
    \centering
    \includegraphics[width=0.3\textwidth]{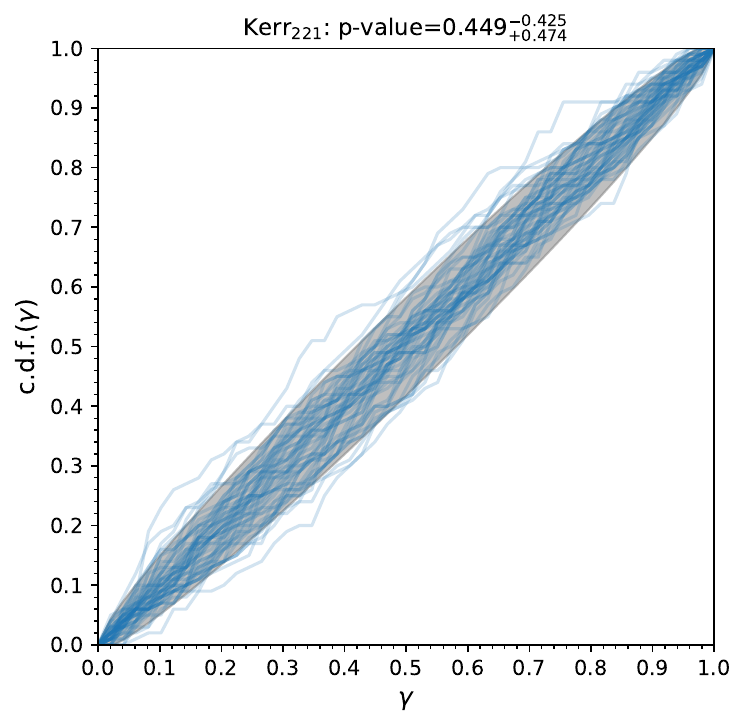}\\
    \includegraphics[width=0.3\textwidth]{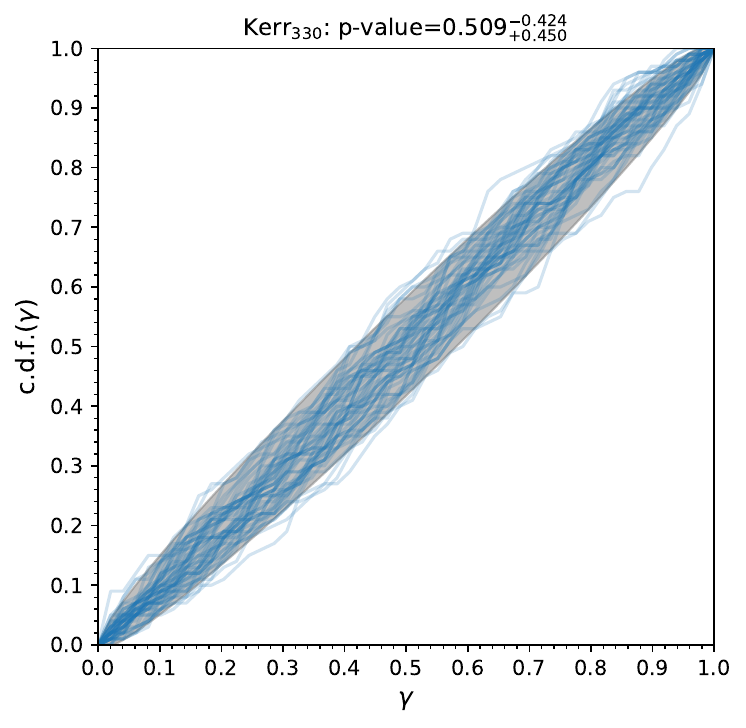}
    \caption{Cumulative distributions of the coverage $\gamma(\boldsymbol{\theta}_*,\boldsymbol{x}_*)$ for the injected models ${\rm Kerr}_{221}$ and ${\rm Kerr}_{330}$. Each blue line corresponds to the c.d.f.~of $\gamma$ from $N_s=100$ draws of $\boldsymbol{\theta}_*$. Each panel plots the cumulatives from $100$ experiments. Shaded grey areas denote the $90\%$ uncertainties over $\text{c.d.f.}(\gamma)$. For each panel, we also quote the median and $90\%$ confidence bounds of the KS test p-values across all draws.}
    \label{fig:coverage:2}
\end{figure}

Figure \ref{fig:coverage:2} displays coverage diagnostics for the the injected models corresponding to ${\rm Kerr}_{221}$ and ${\rm Kerr}_{330}$, complementing the diagnostics for the ${\rm Kerr}_{220}$ presented in Figure \ref{fig:coverage:1} in the main text.

\section{Additional corner plots}
\label{sec:corner:plots}
\begin{figure*}[]
    \centering
    \includegraphics[width=0.5\textwidth]{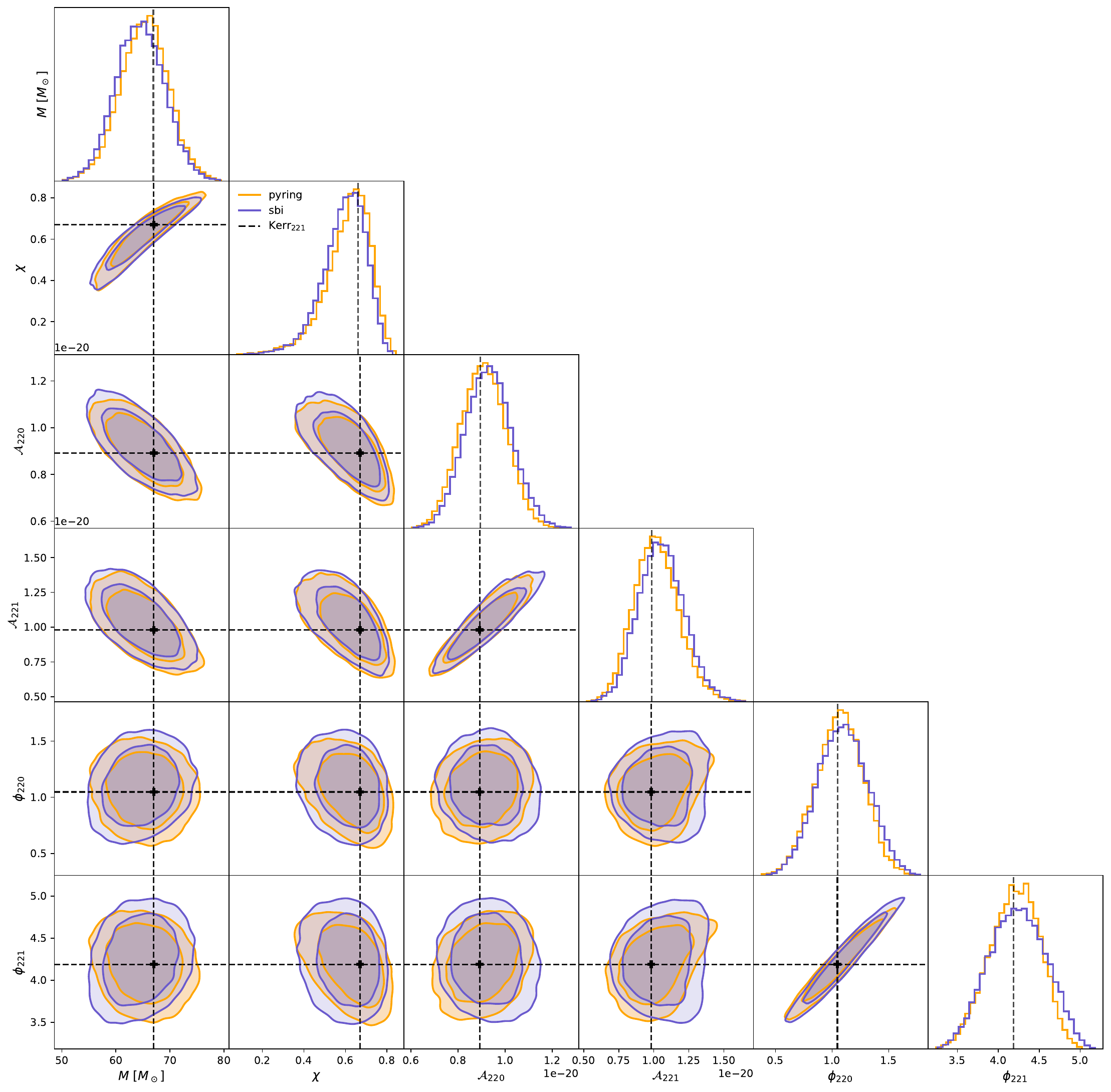}
    \caption{Full corner plot of the posterior recovered from the injected system $\text{Kerr}_{221}$. 2-D contours denote the $68\%$ and $90\%$ credible regions. Black dashed lines denote the injected parameters from Table \ref{tab:injections}.}
    \label{fig:corner:full:221:inj}
\end{figure*}
\begin{figure*}[]
    \centering
    \includegraphics[width=0.55\textwidth]{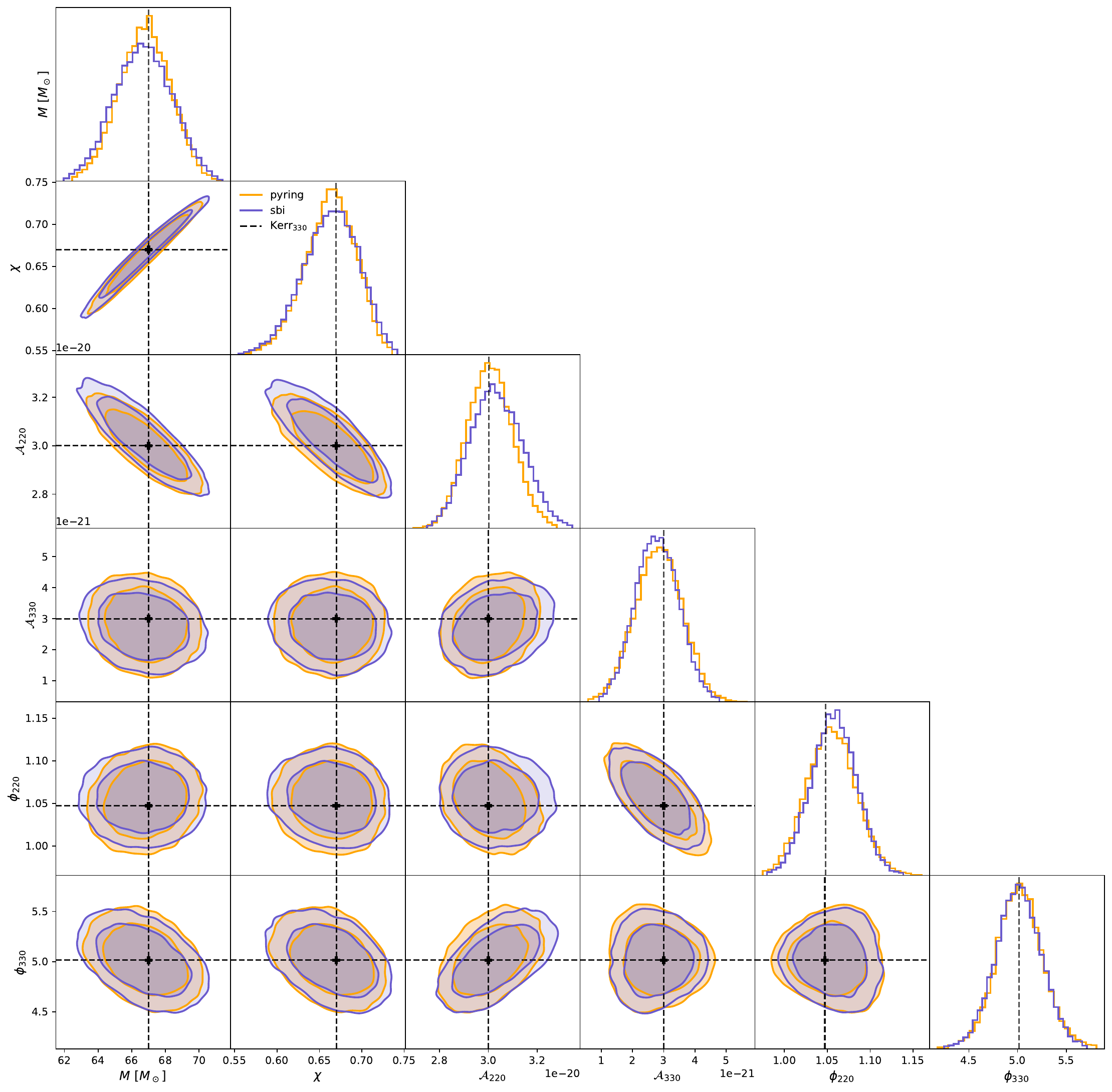}
    \caption{Full corner plot of the posterior recovered from the injected system $\text{Kerr}_{330}$. 2-D contours denote the $68\%$ and $90\%$ credible regions. Black dashed lines denote the injected parameters from Table \ref{tab:injections}.}
    \label{fig:corner:full:330:inj}
\end{figure*}
\clearpage
\bibliography{refs}
\end{document}